\documentclass[9pt,letterpaper,twocolumn]{article}

\usepackage{ol2}
\usepackage[draft]{hyperref}

\usepackage{amsmath}
\usepackage{mathptmx,cite}
\usepackage{graphicx}
\usepackage{amssymb}
\usepackage{amsfonts}
\usepackage{mathrsfs}

\begin{document}

\twocolumn[
\title{Cavity-coupled guided resonances with high quality factors in photonic crystal heterostructures}

\author{M.~Srinivas~Reddy$^{1,*}$ and R. Vijaya,$^{1,2}$}

\address{$^1$Center for Lasers and Photonics, Indian Institute of Technology Kanpur, Kanpur 208016, India\\
$^2$Department of Physics, Indian Institute of Technology Kanpur, Kanpur 208016, India\\
$^*$Corresponding author: msreddy@iitk.ac.in}

\begin{abstract}
We study the optical characteristics of a photonic crystal (PhC) heterostructure cavity consisting of two-dimensional monolayer PhC, sandwiched between two identical passive multilayers. In the range of stopband of the multilayer, guided resonance of the sandwiched PhC are excited by the evanescent waves of the multilayer stack and the quality factor of these cavity-coupled guided resonances is $\sim$$10^6$. The calculated field distribution facilitates the distinction between the cavity defect modes and the coupled guided resonances of the proposed design. The line shapes of the resonances are explained using a theoretical model. Significant decrease in the lasing threshold is observed for these resonant modes in comparison to the defect modes. These results will find use in designing compact PhC-based ultra-low threshold lasers and narrow band filters.
\end{abstract}


]

\noindent Guided resonances in two-dimensional (2-D) photonic crystal (PhC) slabs have prospective applications in the area of optoelectronic devices. These resonances can couple to external radiation and provide efficient channels for the extraction of light from the slabs~\cite{Fan1997,Kanskar1997,Fan2002}.  They have complex dispersion characteristics as well as line shapes and can be excited at $\Gamma$-point of the PhC by the light incident along the normal to the 2-D surface of the structure ~\cite{Fan2002}. These modes with narrow line widths are useful in improving the extraction efficiency of the LEDs, designing low threshold lasing and as narrow band filters in optical communication devices ~\cite{Fan1997,Wiesmann2009,Meier1999,Wang1994}. It is observed that these modes are highly sensitive to the refractive index of the surrounding medium and hence can be used lucratively in sensor applications as well ~\cite{Shi2008}. Major drawbacks of these modes arise from their asymmetrical line shapes and weak vertical confinement ~\cite{Fan2002}.

In this work, we propose a PhC heterostructure cavity consisting of a periodically arranged monolayer of colloids sandwiched between two identical passive multilayers. It is observed that, within the range of the stopband of the multilayer, the evanescent waves can excite the guided resonances of the sandwiched monolayer PhC along with the conventional defect modes. The stopband of the multilayer will provide a strong vertical confinement for the guided mode resonances of the monolayer.

The uniqueness in our proposed cavity arises from the very high quality factor ($\sim$$10^6$) resonances with symmetric line shapes that occur near the center of the stopband of the multilayer. These sharp resonances are useful to design optical devices such as ultra-low threshold lasers and narrow band filters. As an illustrative example for low-threshold lasing, we calculated the lasing threshold characteristics of the proposed structure using Korringa-Kohn-Rostoker (KKR) method ~\cite{Stefanou2000}. The field distribution of these resonant modes in the structure are obtained using DiffractMOD (RSoft$^{TM}$) module. The ideas developed here are equally applicable to any optical device design that benefits from high quality factor.

\begin{figure}[tb!]
\centerline{\includegraphics[width=\columnwidth]{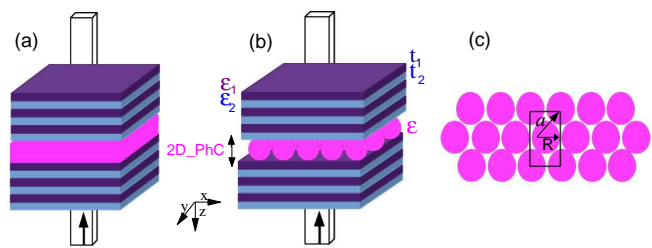}}
\caption{Schematic of the standard multilayer cavity with uniform defect layer (a), and the proposed PhC heterostructure cavity consisting of a monolayer of 2D colloidal PhC with hexagonal lattice sandwiched between the two multilayer stacks (b). In (c), the hexagonal arrangement of colloids in the sandwiched monolayer of 2D PhC is shown. It is assumed to be made up of polystyrene spheres ordered in air background. The rectangle in (c) shows the cross sectional view of the unit-cell used in the DiffractMOD simulation. Cuboids in (a) and (b) show the side view of the unit-cell and vertical arrow shows the direction of incident light.}\label{Fig01}
\end{figure}

Figures ~\ref{Fig01}(a) and ~\ref{Fig01}(b) depict the schematic of the photonic cavity structures used in the calculations. Fig.~\ref{Fig01}(a) is a conventional multilayer cavity having a uniform defect layer. Fig. ~\ref{Fig01}(b) shows the proposed PhC heterostructure cavity. In this design, the uniform defect layer is replaced by a monolayer of 2-D PhC made up of spherical dielectric atoms ordered in a closely packed hexagonal lattice arrangement in air background. The cuboids show the unit cell used in the calculations. The vertical arrow in Fig. 1 shows the direction of incident light. Fig.~\ref{Fig01}(c) shows the hexagonal arrangement in the sandwiched monolayer and the black rectangle in it shows the size of the unit cell used in DiffractMOD.

It is assumed that the monolayer 2-D PhC is made up of polystyrene colloidal spheres with radius (R) and dielectric constant $\varepsilon=2.53$ arranged in closely packed hexagonal lattice with lattice constant `\emph{a}' while the multilayers are composed of alternating $TiO_2$ and $SiO_2$ layers with dielectric constants of $\varepsilon_1=7.02$ and $\varepsilon_2=2.37$ in the wavelength range of interest, and thicknesses of $t_1=0.22a$ and $t_\mathrm{2}=~0.14a$ respectively. We chose the parameters of the multilayer in such a way that it has a broad stopband covering the range of normalized frequencies ($\omega a/2\pi c$) from 0.5 to 1, because the guided mode resonances of the monolayer PhC at $\Gamma$-point will be in this range ~\cite{Kurokawa2002} as shown in the band diagram of the monolayer [see Fig.~\ref{suppl}A in Appendix]. Here c is the speed of light in vacuum. The upper limit of the normalized frequency, namely $\omega a/2\pi c < 1$, is chosen so that the incident light will have less diffraction effects in monolayer PhC ~\cite{Kurokawa2004}.

\begin{figure}[tb!]
\centerline{\includegraphics[width=6cm]{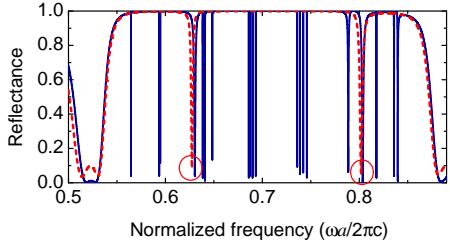}}
\caption{Reflection spectra of the PhC heterostructure cavity [(solid line), for the structure shown in Fig.~\ref{Fig01}(b)] and the multilayer cavity with a defect layer [(dashed line), for the structure shown in Fig.~\ref{Fig01}(a)] having thickness and refractive index equivalent to that of the thickness and effective refractive index of the monolayer PhC. The circles highlight the overlapping modes of the PhC heterostructure cavity and multilayer cavity.}\label{Fig02}
\end{figure}

Figure ~\ref{Fig02} shows the reflection spectra, calculated using KKR method, for the structures shown in Fig.~\ref{Fig01}(a) and Fig.~\ref{Fig01}(b). In calculating the reflectance (dashed curve) for the multilayer cavity with uniform defect layer shown in Fig.~\ref{Fig01}(a), it is assumed that the thickness of the defect layer is equal to the thickness of the monolayer (diameter of the colloid), while its dielectric constant is equal to the effective dielectric constant of the closely packed monolayer. Two defect modes arise due to the defect layer and are separated by the frequency equivalent to $c / 2 n_{eff} a$, where $n_{eff}$ is the effective refractive index of the defect layer. The solid curve in Fig.~\ref{Fig02} is the reflection spectrum calculated for the PhC heterostructure cavity of Fig.~\ref{Fig01}(b), wherein several sharp dips in reflection are seen. The frequencies of the modes of the PhC heterostructure cavity shown by circles are almost matched with the frequencies of the defect modes of multilayer cavity with a uniform defect layer.

When the guided resonances of the monolayer 2-D PhC are superimposed on the flat stopband of the multilayer structure, several sharp resonances are observed in the case of the heterostructure cavity. These sharp resonances are absent in the standard multilayer cavity structure (Fig.~\ref{Fig01}(a)) as well as in the PhC heterostructure cavity when the stopband of the multilayer does not overlap with the guided mode resonant frequencies of the monolayer (see Fig.~\ref{suppl}B in Appendix). These sharp resonant modes are likely to be the guided resonances in the frequency range of 0.5 to 1 (see Fig.~\ref{suppl}A in Appendix) of the monolayer PhC, which are excited by the evanescent field of the multilayers. To understand the characteristics of these sharp resonant modes, we calculated the electric field distribution of the $E_x$ component inside the structure using the DiffractMOD module.

\begin{figure}[tb!]
\centerline{\includegraphics[width=6cm]{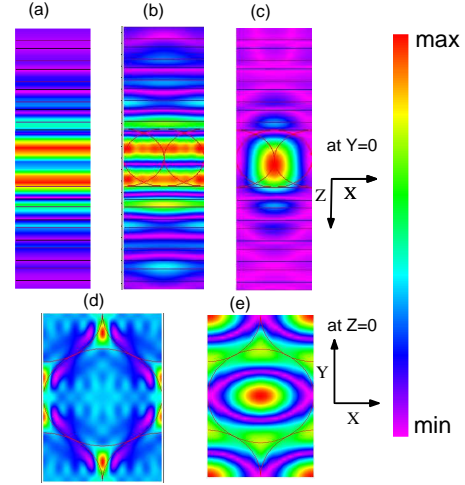}}
\caption{Cross sectional view of the electric field distribution ($E_x$) in the XZ plane at Y=0 of (a) mode with frequency of 0.62732 of multilayer cavity with uniform defect layer (dashed line in Fig.~\ref{Fig02}(a)), mode with frequency (b) 0.63022 (shown as left circle in Fig.~\ref{Fig02}(a)), and (c) 0.69321 of PhC heterostructure cavity. (d), (e) Electric field distributions in the XY-plane at Z=0 for the modes shown in (b), (c). In these calculations, the centre of the unit cell is taken as the origin of the structure.}\label{Fig03}
\end{figure}

The calculated electric field distribution of the $E_x$ component in the cavity structure of Fig.~\ref{Fig01}(a) and ~\ref{Fig01}(b) are shown in Fig.~\ref{Fig03}. Specifically, Fig.~\ref{Fig03}(a) shows the field distribution of the defect mode with frequency 0.627325, in the XZ-plane at Y=0 of multilayer cavity with uniform defect layer. Fig.~\ref{Fig03}(b) and ~\ref{Fig03}(c) show the electric field distribution in the same plane for the modes with frequencies 0.630224 and 0.693216 (see Fig.~\ref{Fig02}) of the PhC heterostructure cavity. It is to be noted that the center of the unit-cell is taken as the origin in these calculations and the periodic boundary conditions are applied along X, Y-directions.

One can observe that the field distributions of Fig.~\ref{Fig03}(a) and ~\ref{Fig03}(b) are similar in nature. The mode with a frequency of 0.807021 of the PhC heterostructure cavity also has a similar field distribution (not shown here). Hence one can conclude that these modes are conventional defect modes of the PhC heterostrure cavity. On the other hand, for the mode with frequency 0.693216 of the PhC heterostructure cavity, the electric field is strongly confined in the dielectric spheres of the monolayer. Other modes (excluding the defect modes with frequencies 0.630224, 0.807021) of the PhC heterostructure cavity also have similar field distributions, wherein the maximum electric field is concentrated in the spheres of the monolayer. The cross-sectional view of the field distributions in the plane passing through the center of the spheres of the monolayer for the defect mode (Fig.~\ref{Fig03}(d)) and the guided mode resonance (Fig.~\ref{Fig03}(e)) also confirm that the electric field of the resonant mode is confined in the sphere while the defect mode is more delocalized. As the guided resonance always shows similar field distributions ~\cite{Kurokawa2004,Lopez2012} in the dielectric spheres of the monolayer as seen in Fig.~\ref{Fig03}(c,e), it confirms that these modes are cavity-coupled guided mode resonances and originate from a process different from that of conventional defect modes.

\begin{figure}[tb!]
\centerline{\includegraphics[width=8cm]{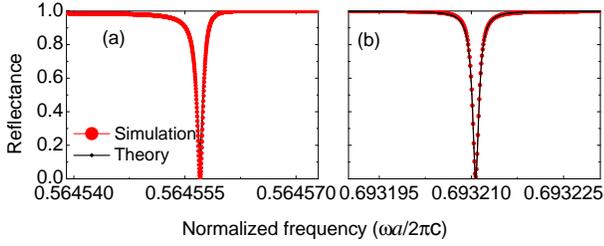}}
\caption{Comparison of theoretical (Eq.~\ref{1}) and numerical line shapes. The solid circles in (a), (b) depict the numerically obtained spectrum using KKR method (see Fig.~\ref{Fig02}). The solid curves are theoretically obtained spectra using Eq.~\ref{1}. The parameters used in theory for these two spectra are (a) $\omega_0=0.564556$, $\gamma=4\times10^{-7}$, and (b) $\omega_0=0.69321$, $\gamma=6.3\times10^{-7}$, which are obtained through numerical method.    }\label{Fig04}
\end{figure}

It is observed from the earlier work ~\cite{Fan2002}, that the lineshape of the stand-alone guided mode resonance of the 2D PhC is asymmetric in nature as a result of interference between the broad background and the leaky modes. These resonances are known as Fano resonances and the reflected amplitude $r$ as a result of the interference between these two waves can be given by  ~\cite{Fan2002}

\begin{equation}\label{1}
  r=r_d\pm f \frac{\gamma}{i(\omega - \omega_0) + \gamma}
\end{equation}

Here $r_d$, $t_d$ are the reflection and transmission co-efficients of the background structure and $\omega_0$, $\gamma$ are the center frequency and the width of the resonance. $f$ is the complex amplitude given by $f= -(r_d \pm t_d)$. In our case, the reflection spectrum of the multilayer cavity with uniform defect layer (dashed line in Fig.~\ref{Fig02}) is providing the broad background.

Our numerical results are compared with the spectrum calculated using Eq.~\ref{1} for two resonant modes (from Fig.~\ref{Fig02}) of PhC heterostructure and are shown in Fig. 4. For the modes near the shoulder of the stopband of the multilayer (for example, the mode with frequency 0.564556) $r_d\neq 1$ and the  line shape of the resonance will be asymmetric as seen in Fig.~\ref{Fig04}(a). On the contrary, $r_d=1, t_d=0$ and the resonance has a symmetric line shape as observed from Fig.~\ref{Fig04}(b) for the modes near the center of the stopband of the multilayer. In the spectra calculated using Eq.~\ref{1}, the values of $\omega_0$, $\gamma$ are obtained numerically using KKR method. One can clearly see a good agreement between the lineshapes obtained by these two methods.

The quality factors of these cavity-coupled Fano resonances are in the range of $\sim$$10^6$. The quality factor of the well designed defect mode (single mode at the center of the stopband) of the structure shown in Fig.~\ref{Fig01}(a) with a cavity length of $\lambda_c/2$ is in the range of $\sim$$10^3$ (see Appendix Fig.~\ref{suppl}C). Here $\lambda_c$ is the center wavelength of the stopband. The enhanced quality factor implies that the proposed heterostructure cavity is a very good candidate for designing nanophotonic laser cavities as well as for use as narrow pass-band filters in comparison to the standard Fabry-Perot cavity with uniform defect layer. To further confirm this, as an example, we calculated the lasing threshold characteristics of these guided mode resonances.

One can expect a significant decrease in the lasing threshold for these guided mode resonances possessing high quality factors as compared to the defect modes due to: (a) their small group velocities, which result from the flat dispersion characteristics ~\cite{Sakoda1999}, and (b) their ability to store their energy in the dielectric spheres containing the gain, which enhances the light-matter interaction. The lasing threshold of these modes is calculated using KKR method ~\cite{Stefanou2000} by incorporating a complex dielectric constant with negative imaginary part for the colloids, which accounts for the gain.

The emission from these cavity structures can be modeled using the complex dielectric constant $\varepsilon=\varepsilon'+i\varepsilon''$, $(\varepsilon''<0)$ ~\cite{Sakoda1999,Sakoda1999a}. Here the negative imaginary part $(\varepsilon'')$ will account for the gain. It is assumed that the gain medium is uniformly doped in the dielectric material (spherical colloids in the monolayer) and the population inversion has been achieved by either optical or electrical pumping and the system is ready to emit. Such colloids are commercially available and have been used by us in our earlier work ~\cite{Reddy2012}. The pumping direction is assumed to be normal to the structure (shown by vertical arrow in Fig.~\ref{Fig01}) and the pumping wavelength is outside the stopband range of multilayer. To obtain the lasing threshold, one can calculate the reflection/transmission as a function of frequency, and $\varepsilon''$. The value of $\varepsilon''$, at which the reflectance/transmittance is divergent, is called as lasing threshold ($\varepsilon''_\mathrm{th}$) ~\cite{Sakoda1999}. For simplicity, we neglected the spontaneous emission in the cavity, which may result in slight overestimation of lasing threshold.

\begin{figure}[tb!]
\centerline{\includegraphics[width=8cm]{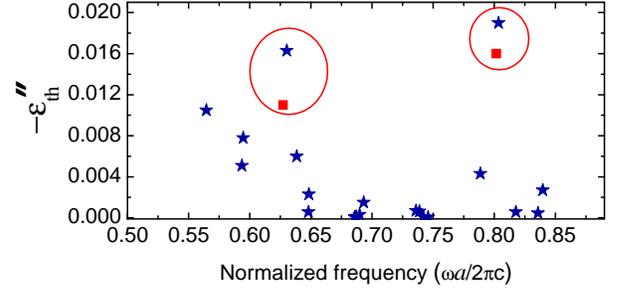}}
\caption{Lasing threshold ($\varepsilon''_\mathrm{th}$) for the modes of the PhC heterostructure cavity (stars) and the defect modes of the multilayer cavity containing a  uniform defect layer (solid squares) (see Fig.~\ref{Fig02}). The circles emphasize the difference in lasing threshold values for defect modes of the PhC heterostructure and multilayer cavity.}\label{Fig05}
\end{figure}

The calculated values of lasing threshold for the modes in Fig.~\ref{Fig02} are shown in Fig.~\ref{Fig05}. The region shown by circles contain the lasing threshold values for modes originating from the defect of the structure. Solid squares are the lasing threshold values for the multilayer cavity with uniform defect layer [shown in Fig.~\ref{Fig01}(a)], but with the defect layer doped to provide gain. It can be seen from Fig.~\ref{Fig05} that the lasing threshold values for the guided resonance frequencies of the PhC heterostructure cavity are significantly lower than that of defect modes. The lasing thresholds for the guided mode resonance near the center of the stopband are lowered by  three orders of magnitude as compared to defect modes. The resonant modes near the shoulders of the stopband have slightly higher values for $\varepsilon''_\mathrm{th}$ in comparison to the resonance near the center of stopband which may be a result of weak vertical confinement. It is important to mention that the guided resonances near the shoulders of the stopband, although have weaker confinement as compared to the defect modes (circles in Fig.~\ref{Fig02}), still show lower threshold values. It is also observed from Fig.~\ref{Fig05} that the lasing threshold for the defect modes of the multilayer cavity with uniform defect layer is less than that of the defect modes of the PhC heterostructure cavity. It is due to the large volume of gain medium in the case of a uniform defect layer whereas in the sandwiched monolayer, the gain medium is assumed to be present only in spheres and not in the air background. The lasing thresholds for the guided mode resonance near the high-frequency shoulder of the stopband have lower values than that near the low-frequency shoulder. It is due to the group velocities of the higher order modes of the PhC being lower than that of low-frequency modes and thus having prolonged interaction with gain medium ~\cite{Sakoda1999a}.

The proposed structure can be easily fabricated using low-cost techniques. The monolayer can be fabricated using self-assembly methods ~\cite{Zhang2012} and multilayer can be fabricated using dip-coating techniques or sputtering ~\cite{Wang1998}. Although we used the monolayer 2-D PhC as a defect in our design, these conclusions will hold for all 2-D PhC slabs with different lattice structures. Moreover, by using high-index materials such as silicon 2-D PhC slabs with air holes as a sandwiched defect layer, one can expect further narrowing in resonance lineshape and thus may result in reducing the lasing threshold further.

In conclusion, we proposed a PhC heterostructure cavity, which supports guided mode resonances of the sandwiched 2-D PhC along with its conventional defect modes. These cavity-coupled guided mode resonances possess high quality factors ($\sim$$10^6$) with symmetrical lineshape functions near the stopband center of the multilayer. The calculated electric field distributions show that the excited guided mode resonance in the proposed structure stores its maximum electric field amplitude in the dielectric sphere and thus enhances the light matter interaction. It is observed that the lasing threshold values of the guided mode resonances are significantly lower than that of standard defect modes. The obtained results are useful in designing PhC-based optical devices such as low-threshold lasers and narrow band filters.

\textbf{Funding.} MSR and RV acknowledge the funds from sponsored project no. GITA/DST/TWN/P-61/2014 that supported the fellowship of MSR in 2014-2015. The work of RV was supported by the Instrument Research and Development Establishment, Dehradun, India under the DRDO Nanophotonics program (ST- 12/IRD-124).

\textbf{Acknowledgment.} RV acknowledges the Director of IRDE for granting the permission to publish this work.

\subsection*{Appendix}
\begin{figure}[h]
\centerline{\includegraphics[width=8cm]{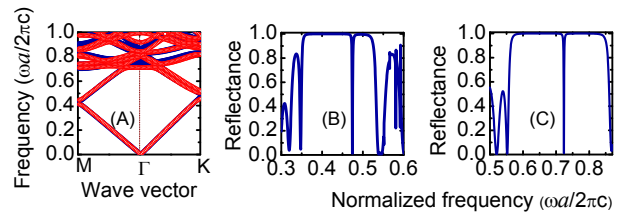}}
\caption{(A) Band diagram for the odd modes (open circles) and even modes (filled circles) of the monolayer 2-D PhC made up of polystyrene colloids arranged in hexagonal lattice. (B) Reflection spectrum of the structure shown in Fig.~\ref{Fig01}(b). Here the thicknesses of the individual layers in multilayer stack are chosen such that their stopband does not overlap with guided mode resonant frequencies of the sandwiched monolayer. (C) Reflection spectrum of the structure shown in Fig.~\ref{Fig01}(a) with defect layer thickness $\lambda_c/2$.}\label{suppl}
\end{figure}
Figure ~\ref{suppl}A shows the band diagram of the monolayer PhC with hexagonal lattice arrangement calculated using plane wave expansion method. The modes at $\Gamma$ point of the PhC can be excited with light incident normal to the surface of the PhC and they will have zero in-plane wave vector. These modes at $\Gamma$ point will fall above the light line and leaky in nature which will couple to the external radiation. Fig.~\ref{suppl}B shows the reflection spectrum of the PhC heterostructure cavity (see Fig.~\ref{Fig01}(b)) with monolayer 2-D PhC as a defect layer. In this calculation, the thicknesses of the individual layers of the multilayer are chosen in such a way that the stopband of the multilayer $(0.35<\frac{\omega a}{2\pi c}<0.54)$ does not overlap with the guided mode resonant frequencies of the sandwiched monolayer PhC. As  a result of this, it can be seen that there is no sharp resonance feature apart from a single defect mode in the spectrum, as in Fig.~\ref{Fig02} in the stopband frequency range. It is to be noted that there are a few sharp asymmetric guided mode resonances existing in the spectrum at frequencies higher than the high-frequency edge of the stopband (see Fig.~\ref{suppl}B, in frequency range 0.55 to 0.6). The asymmetric resonances are a result of interference between the background modes of the multilayer and the guided mode resonances of the 2-D monolayer PhC~\cite{Fan2002}. Fig.~\ref{suppl}C shows the reflection spectrum of the multilayer cavity with uniform defect layer with thickness $\lambda_c/2$. Here $\lambda_c$  is the center wavelength of the stopband. The quality factor of this mode, which is positioned at the center of the stopband is $\sim$700.

\end{document}